\documentclass[12pt]{article}
\usepackage{epsfig,latexsym}

\def\ofo{ { {}_2 \! F_1 }}

\newcommand{\beq}{\begin{equation}}
\newcommand{\eeq}[1]{\label{#1}\end{equation}}
\newcommand{\bea}{\begin{eqnarray}}
\newcommand{\eea}[1]{\label{#1}\end{eqnarray}}

\input epsf
\input{epsf.sty}

 \csname
@addtoreset\endcsname{equation}{section}
       
\begin{document}

\font\cmss=cmss10 \font\cmsss=cmss10 at 7pt 

\hfill NYU-TH/01/04/01
 \vskip .1in\hfill SNS-PH/01-04   
\vskip .1in \hfill hep-th/0104065

\thispagestyle{empty}

\hfill

\vspace{20pt}

\begin{center}
{\Large \textbf{\em On the Absorption by Near-Extremal Black Branes}}
\end{center}

\vspace{6pt}

\begin{center}
\textsl{ G.~Policastro$\,^{a,b}$ and A.~Starinets$\,^a$ } \vspace{20pt}

\textit{$(a)$ Department of Physics, New York University,\\
 4 Washington Place, New York NY 10003 USA \\ \vspace{.2cm}
$(b)$ Scuola Normale Superiore, \\
Piazza dei Cavalieri  7, 56100, Pisa, Italy}


\end{center}

\vspace{12pt}

\normalsize

\begin{center}
\textbf{Abstract }
\end{center}

\vspace{4pt} {\small \noindent 

We study the absorption of a minimally coupled scalar in the
gravitational background created by a stack of near-extremal black
three-branes, and more generally by $M2$, $M5$ and $Dp$ branes.  The
absorption probability has the form $P(l) = P_0(l) f_l(\lambda )$,
where $P_0(l)$ is the partial wave's absorption probability in the
extremal case, and the thermal factor $f_l(\lambda )$ depends on the
ratio of the frequency of the incoming wave and the Hawking
temperature, $\lambda = \omega/\pi T$. Using Langer-Olver's method, we
obtain a low-temperature ($\lambda \gg 1$) asymptotic expansion for
$P(l)$ with coefficients determined recursively. This expansion, which
turns out to be a fairly good approximation even for $\lambda \sim 1$,
accounts for all power-like finite-temperature corrections to
$P_0(l)$, and we calculate a few terms explicitly. We also show that
at low temperature the absorption probability contains exponentially
suppressed terms, and attempt to develop an approximation scheme to
calculate those. The high-temperature expansion is also
considered. For the $s$-wave, the low-temperature gravity result is
consistent with the free finite-temperature field theory calculation,
while for high temperature and higher partial waves we find a
disagreement.  As a check of the approximation methods used, we apply
them to the $D1-D5$-brane system, and compare results to the known
exact solution.}

\eject
\noindent

\section{Introduction}\label{intro}

It is believed that a strongly coupled 
gauge theory at finite temperature 
can be studied using its gravity dual 
description in the framework
 of the AdS/CFT correspondence.
Examples relevant for $d=4$ gauge theories include Witten's interpretation
of the confinement-deconfinement phase transition in terms of
 the Hawking - Page transition in AdS and 
calculation of the free energy of the 
finite-temperature  ${\cal N}=4$
SYM which matches (up to a factor of 4/3)
 the free energy of the near-extremal 
three-branes  (for review and references see
section 3.6 of \cite{Aharony:2000ti}).

Calculation of the absorption cross section by $D3$-branes revealed 
a remarkable agreement between conformal field theory
 and gravity results, obtained, correspondingly, at weak and 
strong 't Hooft coupling \cite{Klebanov:1997kc}.
 The agreement has been explained (at least for
the $s$-wave absorption) by establishing a non-renormalization theorem
enjoyed by certain ${\cal N}=4$ SYM correlation functions 
\cite{Gubser:1997se}, and serves as
one of the few direct checks of the conjectured AdS/CFT correspondence.
Supersymmetry of the $D3$-branes' world-volume theory is crucial for
the existence of the above mentioned non-renormalization theorem.
At finite temperature the supersymmetry is broken, and  one cannot expect
the results computed at strong and weak coupling to agree. 

However, for the $D1/D5$ model of a five-dimensional black hole the
agreement between greybody factors and the entropy 
calculated in two-dimensional CFT and 
in gravity holds even in the absence of supersymmetry.
The absorption cross-section has the form
$\sigma = \sigma_0 f(\omega , T_L, T_R)$,
where $\sigma_0$ is the cross-section at zero temperature and
the thermal factor $f(\omega , T_L, T_R)$ can be computed
either from (super)gravity 
(in the low energy regime, i.e. when $\omega r_1 \ll 1$,
 $\omega r_5 \ll 1$) or by considering a finite-temperature
two-point function in the dual CFT
\cite{Das:1996jy}, \cite{Das:1996wn}, \cite{Maldacena:1997ih}. 
A perfect matching between CFT and gravity results can be explained by 
arguing that at sufficiently 
low energies the excitations of the $D1/D5$ system
are correctly described by the moduli space approximation even 
at strong coupling \cite{Maldacena:1997iz}.

{\it A priori}, the existence of a non-renormalization theorem not
related to supersymmetry in $d=4$  can be considered as an extremely
remote possibility, and indeed in this paper 
we do not find an exact agreement between
gravity and weakly coupled field theory calculations.
However, the details of the {\it disagreement} are intriguing enough,
and in the spirit of AdS/CFT we regard our results obtained from
gravity as a prediction for a strongly coupled finite-temperature 
gauge theory.

For the  D1-D5-brane system, it is fairly straightforward to calculate 
the thermal factor
$f(\omega , T_L, T_R)$ on the gravity side,
since in the low energy regime the exact solution of the corresponding
 wave equation is known.
 This is not the case for the generic black brane
solution, and for the three-brane solution corresponding to 
finite temperature $d=4$ gauge theory in particular.
It is a purpose of this paper to develop a reliable approximation scheme
which would allow one to calculate greybody factors for the case 
of a non-extremal three brane, and more generally for $M2$-, $M5$- and
$Dp$-branes.
Earlier attempts to calculate the absorption probability for
non-extremal three-branes were made in \cite{Satoh:1998ss}, 
\cite{Musiri:2001dj}, \cite{Vazquez-Poritz:2000ms}. 
However, we will show that only the
 zero-temperature limit of the results
obtained there is correct.

In the low energy regime ($\omega R\ll 1$) 
the radial part of the wave equation in a typical black-brane background
has several regular singular points. Three of them always correspond, respectively, to the position of outer, inner horizons and spatial infinity, and the rest lies on a unit circle in the complex plane.
 Since the number of singular points exceeds three,
 the exact solution is unaccessible. However, the equation contains a parameter $\lambda = \omega/\pi T$ making it possible to 
study ``low-temperature'' ($\lambda \gg 1$) and ``high-temperature''
 ($\lambda \ll 1$) asymptotics.
While obtaining the high-temperature asymptotics is a straightforward application of perturbation theory, the low-temperature expansion requires 
a careful WKB analysis of the differential equation in the presence of singular points.
Basic principles of such an analysis have been developed a long time ago 
in
\cite{langer, olver_1} (see also \cite{swanson}
 and a classic book by Olver \cite{Olver_book}). 
Using Langer - Olver's method, we are able to obtain 
uniform asymptotic expansions 
(with coefficients determined recursively) for the solutions of the 
wave equation. The uniformity of approximation is vital 
for the correct calculation of the absorption probability.
Olver's theorem guarantees this property
only when the coefficients and the corresponding error-control
function converge everywhere on the interval of consideration.
For non-extremal branes this interval always contains two
singular points
(a situation more general than the one
 originally treated 
in  \cite{langer, olver_1}), and therefore the 
convergence should be checked in each case independently. 
Once the convergence has been established, all power-like corrections to 
the zero-temperature (extremal) result can in principle be determined
using the recursive relations.
We obtain explicit asymptotic expansions for the black three-brane
background (Section \ref{D3_Langer}) and for 
non-extremal $D1-D5$-brane system (Section \ref{D1D5}), 
and demonstrate (in Section \ref{M_branes}) 
that all $M$- and black-brane backgrounds can be treated
using this approach.
Being an asymptotic expansion in powers of $1/\lambda$, the method cannot
account for the contributions of the form
$\exp{(-\lambda)}$
(and for the $s$-wave those are the only ones present).
In Section \ref{D3_hyper}
 we make an attempt to build an approximation scheme
based on the hypergeometric function which would allow the calculation of 
these
exponentially small contributions. We find the leading term of the corresponding expansion, but the subleading corrections are controlled by an integral equation which we were unable to solve.
Section \ref{D3_high_T} deals with the high-temperature expansion.
In Section \ref{field_theory} we compute the absorption cross section in 
field theory using the
method of finite-temperature Green's functions. 
Appendices \ref{appendix_a}, \ref{appendix_b}, \ref{appendix_d}
 contain some relevant technical details.

\section{Summary of results} \label{summary}

\noindent
Our main result can be formulated as follows: the absorption
 probability 
 for a minimally coupled 
massless  scalar in
the gravitational background created by a stack of 
near-extremal black three-branes can be written as
$P(l) = P_0(l) f_l(\lambda)$, where
\begin{equation}
P_0 (l) = \frac{4\pi^2 }{(l+1)!^2 (l+2)!^2}
 \left( \frac{\omega R}{2}\right)^{4l+8}
\label{zero_temp_abs}
\end{equation}
is the absorption probability for the extremal $D3$-branes. 
The absorption cross section (greybody factor) 
for the $l$-th partial wave is related to the absorption 
probability $P_l$ via
\begin{equation}
\sigma_l = \frac{2\pi^{\frac{d-1}{2}}
\Gamma \left( l+d-2 \right) \left( l+d/2 - 1\right)}{
\omega^{d-1}\Gamma \left( d/2 -1/2\right) \Gamma \left( l+1 \right)} P(l)\,,
\end{equation}
where $d$ is the number of dimensions transverse to the brane
\cite{Gubser:1997qr}.
For the thermal factor $f_l(\lambda)$ we obtain a low-temperature 
 ($\lambda = \omega/\pi T \gg1$)  asymptotic expansion
\begin{eqnarray}
 f_l(\lambda) &=&
 1 + \frac{2\nu (\nu^2-1)(\nu^2-4)}{5\lambda^4} \nonumber \\
&+& \frac{2\nu (\nu^2-1)(\nu^2 - 4)(\nu - 3) 
(9\nu^4 -23 \nu^3 - 114\nu^2 +188\nu +600)}{225\lambda^8} \nonumber \\ &+&
O\left( 1/\lambda^{12}\right)\,,
\label{d3_result}
\end{eqnarray}
where $\nu = l+2$.
The function  $f_l(\lambda)$ for $\lambda \gg 1$ also contains
exponentially small terms (see (\ref{absorption_hyper})).
 The high-temperature limit
of  $f_l(\lambda)$ for various $l$ 
is given by (\ref{high_T_odd}), (\ref{high_T_even_0}),
(\ref{high_T_even_2}).
In particular, for the $s$-wave the gravity calculation gives
\begin{equation}
   f_{0, \, \mbox{\tiny grav}}(\lambda )   \;=\;
   \cases{   1 + O \left( e^{-\lambda}\right)
                                                   & for $\lambda \gg 1$,  \cr
           \noalign{\vskip 4pt}
          \displaystyle{ \frac{8}{\pi \lambda^3}\left( 1 +  \frac{
\left(\pi^2+12(2-\log{2})\log{2}\right)\lambda^2}{48}
 + \cdots \right)}
                                                   & for $\lambda \ll 1$,   \cr
         }
   \label{f_grav}
\end{equation}
while in field theory we get  $f_{0, \, \mbox{\tiny field th.}}(\lambda ) =
\coth{(\pi\lambda /4)}$, i.e.
\begin{equation}
   f_{0, \, \mbox{\tiny field th.}}(\lambda )   \;=\;
   \cases{   1 + O \left( e^{-\lambda}\right)
                                                   & for $\lambda \gg 1$,   \cr
           \noalign{\vskip 4pt}
          \displaystyle{
 \frac{4}{\pi \lambda}\left( 1 +  \frac{\pi^2\lambda^2}{48}
 + \cdots \right)}
                                                   & for $\lambda \ll 1$.   \cr
         }
   \label{f_fth}
\end{equation}
For $l=1$ the disagreement is visible already at low temperature,
\begin{eqnarray}
 f_{0, \, \mbox{\tiny grav}}(\lambda ) &=& 1 + 48/\lambda^4 + O\left( e^{-\lambda} \right)\,,\label{f_l_1_grav} \\
 f_{0, \,\mbox{\tiny field th.}}(\lambda ) &=& 1 + 8/\lambda^2 +  48/\lambda^4 + 
 32/\lambda^6 + 
O\left( e^{-\lambda} \right)\,.
\label{f_l_1_field}
\end{eqnarray}
As mentioned above, we do not expect free field theory and gravity results to 
agree, especially in the regime $\omega \ll T$, where the resummation of an 
infinite subset of diagrams in the finite-temperature gauge theory is required.
We rather regard (\ref{f_grav}) as a prediction for a strongly coupled 
gauge theory at non-zero temperature.

\section{Absorption by non-extremal black three-branes}\label{D3}

\noindent
The metric of the black three-brane is given by \cite{Horowitz:1991cd}
\begin{equation}
ds^2_{10} = \frac{-f(r)dt^2 + d\vec{x}\cdot d\vec{x}}{\sqrt{H(r)}}
 +\sqrt{H(r)}\left( \frac{dr^2}{f(r)}+r^2 d\Omega_5^2\right)\,,
\label{metric} 
\end{equation}
where 
$$
H(r) = 1 + \frac{R^4}{r^4}\,, \;\;\;\;  f(r) =  1 - \frac{r_0^4}{r^4}\,.
$$
The position of the inner horizon 
corresponds to $r=0$, the outer horizon is located at $r=r_0$. 
We are interested in the region of the parameter space characterized by
 $\omega R \ll 1$, $T R\ll 1$ with the ratio
 $\lambda = \omega /\pi T$ remaining arbitrary.
 The condition  $T R\ll 1$ implies that 
the metric we consider is  near-extremal ($r_0\ll R$).
For $r_0\ll R$ the Hawking temperature
is related to the non-extremality parameter $r_0$ by 
 $T=r_0/\pi R^2$. The horizon area is given
 by $A_H = \pi^3 r_0^5 \sqrt{H(r_0)}
\rightarrow \pi^3r_0^3R^2$ for $r_0\ll R$.

\noindent
Using $\rho = \omega r$,  $\rho_0 = \omega r_0$, one can
 write the radial part of the 
Laplace equation for a minimally coupled massless scalar as
\begin{equation}
\frac{d^2 \phi}{d\rho^2} + \frac{5\rho^4 - \rho_0^4}{\rho(\rho^4 - \rho_0^4)}
\frac{d \phi}{d\rho} - \frac{\rho^2l(l+4)}{\rho^4 - \rho_0^4} \phi +
\frac{\rho^4 (\rho^4 +(\omega R)^4)}{(\rho^4 - \rho_0^4)^2} \phi =0\,.
\label{rho_equation}
\end{equation}
In the extremal case $r_0=0$, equation 
(\ref{rho_equation}) in terms of $z=r^2$ 
reduces to
\begin{equation}
4 z^3 \phi '' + 12 z^2 \phi ' + 
\left( \omega^2 R^2 - l(l+4) z + \omega^2 z^2 \right) \phi = 0 \,,
\end{equation}
solutions of which are known\footnote{See 2.306 of \cite{kamke}.
A complete absorption analysis based on the exact
 solution has been done in 
\cite{Gubser:1999iu}.}.
Exact solutions of (\ref{rho_equation}) are not known, 
and therefore one must seek a reliable approximation scheme.
Introducing $x = r_0^2/r^2$, we can write   (\ref{rho_equation}) as
\begin{equation}
\frac{d^2 \phi}{d x^2}  - \frac{(1+x^2)}{x (1-x^2)}
 \frac{d \phi}{d x} - \frac{l(l+4)}{4 x^2 (1-x^2)} \phi +
\frac{(\lambda^2 + \rho_0^2/x^2)}{4 x (1-x^2)^2} \phi = 0\,.
\label{xx_eq}
\end{equation}
Following \cite{Klebanov:1997kc}
 we shall calculate the absorption probability to
 the leading order in $\omega R \ll 1$ by introducing the 
``inner'' and ``outer'' regions for the equation  (\ref{xx_eq}), 
and matching the corresponding solutions. 
In the 
{\bf{\em outer region}}, $\rho \gg \rho_0$, $\rho \gg (\omega R)^2$,
equation  (\ref{rho_equation}) reduces to a Bessel equation with the 
solutions
\begin{equation}
\phi (r) = \alpha J_{l+2} (\rho )/\rho^2 +  \beta Y_{l+2} (\rho )/\rho^2\,.
\end{equation}
In the {\bf{\em inner region}}, $\rho_0 \leq \rho \ll 1$, 
the last term in (\ref{xx_eq}), 
$\rho_0^2/x^3$ can be neglected in comparison with 
 the typical\footnote{The term with $l$
is an example of such a typical term.
 However, even for $l=0$ the ``potential''
characterizing the geometry is proportional to $1/x^2$ - 
this can be seen by converting the equation into the Schr\"{o}dinger form.}
 ``potential'' term $\sim 1/x^2$.
Equation
 (\ref{xx_eq}) becomes
\begin{equation}
\frac{d^2 \phi}{d x^2}  - \frac{(1+x^2)}{x (1-x^2)}
 \frac{d \phi}{d x} - \frac{l(l+4)}{4 x^2 (1-x^2)} \phi +
\frac{\lambda^2 }{4 x (1-x^2)^2} \phi =0\,.
\label{x_eq}
\end{equation}
 It has four regular singular points (at $x
 = \pm 1, 0, \infty$).
 Unlike the
 case of a second order differential equation with three regular singular
 points which can always be cast into the form of hypergeometric
 equation by performing a suitable conformal transformation, such an
 equation has no common ancestor and should be studied on its own.
 The equation, however, contains a parameter $\lambda$
 making it possible to study low- and high-temperature
 expansions even though the exact solution remains unaccessible.
\noindent
Finally, in the {\bf {\em matching region}},
 $\rho_0 \ll (\omega R)^2 \ll \rho \ll 1$, we have
\begin{equation}
\phi (r) \rightarrow \alpha \rho^l / 2^{l+2} (l+2)!  -
 \beta 2^{l+1}(l+1)!/\pi \rho^{l+4}\,.
\end{equation}
Anticipating that for $\omega R\ll 1$
 $\beta \rightarrow 0$, the 
absorption probability can be written as
\begin{equation}
P(l) = 2\pi \lambda \rho_0^4/ |\alpha |^2
\label{abs_prob}
\end{equation}
%

\noindent
To study solutions of 
(\ref{x_eq}),
it will be convenient to eliminate the first derivative term by introducing
$u(x) = x^{-1/2}\sqrt{x^2-1} \phi (x)$
and rewriting
(\ref{x_eq}) in the form
\begin{equation}
u'' = \left( \lambda^2 F(x) + G(x)\right) u\,,
\label{x_normal_form}
\end{equation}
where 
$$
F(x) = - \frac{1}{4 x (1-x)^2 (1+x)^2}\,, \;\;\;\hspace{0.5cm}
G(x) = - \frac{x^4+6 x^2 -3}{4 x^2 (1-x^2)^2} + \frac{l(l+4)}{4 x^2 (1-x^2)}\,.
$$
In the next subsection,
 we shall find solutions to equation (\ref{x_normal_form})
in the form of asymptotic series in  $1/\lambda$.

\subsection{Low-temperature expansion I: Langer - Olver method}\label{D3_Langer}

\noindent
We are interested in constructing a uniform approximation to the solution
of (\ref{x_normal_form}) on the interval $[0,1]$, the endpoints of 
the interval being regular singular points of the differential equation.
Our expansion will be built around $x=0$,
 where the function $F(x)$ has
 a simple pole\footnote{The expansion cannot be 
constructed starting from $x=1$, since in that case the error-control function
 diverges (see remarks on p.205 of \cite{Olver_book}).}. 
The idea behind the method of Langer and Olver is to
introduce new variables in which the behavior of the solution close
 to the singularity  would be fully 
described (for large $\lambda$) by a known function, 
and treat the rest as a perturbation. In our case the
 situation is more complex since the interval of interest contains {\it two}
singular points, and we have to prove that the functional coefficients 
of the expansion converge everywhere on $[0,1]$ - in particular,
 that they converge at $x=1$. 
The convergence can be established by analyzing the singularities
of the functions $F(x)$ and $G(x)$ in (\ref{x_normal_form}) \cite{Olver_book}.
We also confirm it by an explicit calculation.

\noindent
By making a conformal transformation $\zeta = - p^2(x)$, where 
\begin{equation}
p(x) = \frac{1}{2}\arctan{\sqrt{x}} +\frac{1}{4}
 \log{\frac{1+\sqrt{x}}{1-\sqrt{x}}}
\end{equation}
and introducing a new function  $W(x) = x^{-1/4}\sqrt{p(x)/(x^2-1)} u(x) = 
 x^{-3/4}\sqrt{p(x)} \phi (x)$
we transform equation (\ref{x_eq}) into
\begin{equation}
\frac{d^2 W}{d\zeta^2} = \left\{ \frac{\lambda^2}{4\zeta} + 
\frac{\nu^2-1}{4\zeta^2} + \frac{\psi (\zeta )}{\zeta}\right\} W\,,
\label{W_d3}
\end{equation}
where $\nu=l+2$, and 
\begin{equation}
\psi ( \zeta (x) ) = \frac{4\nu^2 -1 }{16 p^2(x)} + 
\frac{(x^2-1)(9x^2+4\nu^2 - 1)}{16 x}\,.
\end{equation}
The function $\psi (\zeta )/\zeta$ is analytic at $\zeta = 0$ ($x=0$) and
is treated as a perturbation. 
Then the formal series solution of (\ref{W_d3}) corresponding to the 
wave propagating from the infinity to the horizon is given by
\begin{equation}
W(x) = \sqrt{\frac{\pi \lambda}{2}}
\frac{ i^{l+2}p(x)}{ N_1(\lambda ,l)}
\left\{ H^{(1)}_{l+2}(\lambda \,p(x))
\sum\limits_{n=0}^{\infty} A_n(x) \lambda^{-2n} - \frac{p(x)}{\lambda}
 H^{(1)}_{l+3}(\lambda \,p(x))
\sum\limits_{n=0}^{\infty} B_n(x) \lambda^{-2n}\right\}\,,
\label{D3_W}
\end{equation}
where
$H_{\nu}^{(1)}(z)$ are Hankel functions.
The  coefficients $A_n$, $B_n$ are
 independent of $\lambda$ and can be
 calculated  successively using the following equations \cite{olver_1}
\begin{equation}
A_{n+1}(\zeta ) = \nu B_n(\zeta ) - \zeta B'_n(\zeta )
 + \int \psi (\zeta ) B_n(\zeta)d\zeta \,,
\label{a_s}
\end{equation}
\begin{equation}
B_{n}(\zeta ) = -  A'_n(\zeta ) + \frac{1}{\sqrt{\zeta}}
\int\limits_{0}^{\zeta}\left\{ \psi (t) A_n(t) - (\nu +1/2)A'_n(t)\right\}
\frac{dt}{\sqrt{t}} \,,
\label{b_s}
\end{equation}
with $A_0=1$.
The formal series in (\ref{D3_W}) provides a genuine asymptotic
expansion under the condition that the successive integrals in
(\ref{a_s}), (\ref{b_s}) are convergent \footnote{Precise error
 bounds on partial sums of (\ref{D3_W}) 
can be found in Chapter 12, $\S$ 4 of \cite{Olver_book}).
 Conditions on  $\psi$
sufficient (but not necessary!) for 
the convergence of the error-control function are stated in
Exercise 4.2, Chapter 12 of \cite{Olver_book}.
}.
As can be readily seen
\footnote{For $x\rightarrow 1$ 
we have $B_n \sim 1/\log{\epsilon},
 A_n \sim O(1)$, where $\epsilon = |x-1|$.},
 this is indeed the case for our $\psi (x)$, and therefore
Olver's theorem guarantees that 
(\ref{D3_W}) is an asymptotic expansion 
approximating the solution of (\ref{W_d3})  {\it uniformly}
 for $x\in [0,1]$.
 In practice, it is far more convenient to use the following expressions
for the coefficients $A_n(x)$, $B_n(x)$ (see 
 Chapter 12, Exercise 5.2 in \cite{Olver_book}):
\begin{equation}
A_n(x) = \sum\limits_{k=0}^{2n}
\frac{i^k a_k(\nu +1)}{p^k(x)} \bar{A}_{2n-k}(x)\,,
\label{bar_a}
\end{equation}
\begin{equation}
B_n(x) = 
i\sum\limits_{k=0}^{2n+1}
\frac{i^k a_k(\nu )}{p^{k+1}(x)} \bar{A}_{2n+1-k}(x)\,,
\label{bar_b}
\end{equation}
where
\begin{equation}
\bar{A}_{k+1}(x) = 
i\sqrt{x}(x^2 - 1)\partial_x 
\bar{A}_k + i\int \frac{9 x^2 + 4\nu^2 - 1}{16x\sqrt{x}} \bar{A}_k dx\,,
\label{reccur}
\end{equation}
and $a_k(\nu ) = (4\nu^2-1^2)(4\nu^2-3^2)\dots (4\nu^2-(2k-1)^2)/8^k 
k!$.
\noindent
The normalization constant $N_1(\lambda ,l)$ in (\ref{D3_W}),
\begin{equation}
N_1(\lambda ,l) =  e^{i\lambda \log{2}/2 - i\pi /4 +i\lambda \pi /8}
 \sum\limits_{n=0}^{\infty}\left(A_n(1) - \frac{i}{\lambda}
 \lim_{x\rightarrow 1}p(x)B_n(x)\lambda^{-2n}\right)\,,
\end{equation}
is chosen in such a way that for $x\rightarrow 1$ we have
$\phi (x) =
 x^{3/4} W /\sqrt{p(x)} \rightarrow (1-x)^{-i\lambda /4}$.
Then in the limit $x\rightarrow 0$ we obtain 
\begin{equation}
\phi (x) \rightarrow \sqrt{\frac{2\lambda }{\pi}} \frac{i^l (l+1)! 2^{l+1} 
N_0}
{\lambda^{l+2} N_1} x^{-l/2}\,,
\end{equation}
where
\begin{equation}
N_0 (\lambda ,l) = \sum\limits_{n=0}^{\infty}\lambda^{-2n}
\left( A_n(0) - 2\nu B_n(0)\lambda^{-2}\right)\,,
\end{equation}
which can be compared with $\alpha J_{l+2}(\rho )/\rho^2$ in the region $\rho \ll 1$ to get the coefficient $\alpha$. The absorption probability 
is calculated in a standard way, and we find
\begin{equation}
P(l) = P_0 (l) \frac{|N_1(\lambda ,l)|^2}{|N_0(\lambda ,l)|^2}\,,
\end{equation}
where $P_0(l)$ is given by (\ref{zero_temp_abs}).
Calculating the first few coefficients, we get
$B_0(0)=0$, $\lim_{x\rightarrow 1}p(x)B_0(x) = (\nu^2-1)/2$,
$A_1(0) = (\nu^2-1)^2/8$ (see also Appendix A). 
Then the ratio $f_l(\lambda) = P(l)/P_0(l)$ is given by
(\ref{d3_result}).
Higher order terms can be calculated as a result of a lengthy
 but essentially straightforward procedure using the recurrence relations
(\ref{bar_a} - \ref{reccur}). The result (\ref{d3_result}) therefore
potentially accounts for all power-like terms in the $\lambda \gg 1$ 
expansion. From our treatment of the $D1-D5$-brane system in Section 3 it will
become clear that the Langer - Olver expansion approximates the exact
 solution reasonably well even for $\lambda \sim 1$.
In fact, the first few terms of the asymptotic expansion 
(\ref{D3_W}) 
coincide with the Frobenius series solution of (\ref{W_d3}) at $x=0$.

The following assertions regarding the asymptotic expansion 
(\ref{d3_result})
  {\it appear} to be true:
\begin{itemize}
   \item[(1)] It is an expansion  in powers
 of $1/\lambda^4$. In particular, the $1/\lambda^2$
 correction vanishes, in disagreement\footnote{In \cite{Satoh:1998ss},
the potential was approximated by its Taylor expansion 
near one of the singular points. Such an approximation is non-uniform
and cannot give an accurate matching.
The approximation used in
 \cite{Musiri:2001dj}, \cite{Vazquez-Poritz:2000ms}
is essentially of the form $f(x) = x^{\alpha_0}(1-x)^{\alpha_1}$, 
where $\alpha_{0,1}$ are indices of the singularities
 at $x=0,1$, respectively. This approximation provides the correct 
result for $P$ at $\lambda =\infty$ 
(absorption probability at zero temperature) but it cannot be used 
to calculate finite-temperature corrections. We remark that the results of
 \cite{Musiri:2001dj},
\cite{Vazquez-Poritz:2000ms} disagree with those obtained in 
 \cite{Satoh:1998ss}. }
 with  \cite{Satoh:1998ss}, \cite{Musiri:2001dj},
\cite{Vazquez-Poritz:2000ms},
\item[(2)] 
 For the $s$-wave ($\nu = 2$) power-like corrections in  (\ref{d3_result})
vanish.
\end{itemize}
Being an asymptotic expansion in powers of $1/\lambda$, formula
(\ref{d3_result}) cannot account for the 
 corrections of the
 form $\exp{(-\lambda})$. In the next section we show 
that such terms do indeed exist. 
Unfortunately, an efficient recursive scheme 
for their calculation is not known \cite{olver_private}.
 Our attempt to construct it seems to face
 considerable technical difficulties.

\subsection{Low-temperature expansion II: Reduction 
to a hypergeometric equation}\label{D3_hyper}

\noindent
Since we are interested in studying solutions in the interval $[0,1]$,
it seems plausible that singular points $x=0$, $x=1$ play the major role for such solutions, whereas the singular point $x=-1$ can be treated perturbatively.
Indeed, neglecting the term with $G(x)$ for a moment,
we can write  (\ref{x_normal_form}) as
\begin{equation}
u'' = - \frac{\lambda^2 u}{4x(1-x)^2\eta (x)}\,,
\label{crude}
\end{equation}
where $\eta (x) = (1+x)^2$ varies smoothly from $1$ to $4$ for $x\in [0,1]$.
If we treat $\eta$ as a constant, equation (\ref{crude})
 can be solved exactly in terms of a hypergeometric function,
 and therefore one may hope that the solution of the original
 equation (\ref{x_normal_form})
 can be successfully approximated by a series based on
 hypergeometric function. 
We shall now try to make these ideas precise.

First, we introduce a new variable, $\zeta (x)\in [0,1]$, and make a 
conformal transformation,
\begin{eqnarray}
\zeta &=& \tanh^2 p(x)\,, \\
p(x) &=&  \left( \mbox{arctan} \sqrt{x}  + 
\mbox{arctanh} \sqrt{x} \right)/2\,, \\
U &=& u \sqrt{ d\zeta /d x }\,,
\end{eqnarray}
to write equation (\ref{x_normal_form}) in the form
\begin{equation}
U'' = - \frac{\lambda^2}{4\zeta (1-\zeta)^2} U + {\cal P} U\,,
\label{U_eq}
\end{equation}
where derivatives are now taken with respect to $\zeta$ and
\begin{equation}
{\cal P} = \dot{x}^2 g(x(\zeta )) - \frac{1}{2}\left\{ x,\zeta \right\} \,.
\end{equation}
Here
 $\dot{x} = dx/d\zeta = \sqrt{x}(1-x^2)\cosh^2 p(x)/\tanh{p(x)}$, and 
$$
\left\{ x,\zeta\right\} = \frac{\stackrel{\cdot\cdot\cdot}{x}}{\dot{x}} 
- \frac{3}{2}\left( \frac{\ddot{x}}{\dot{x}}\right)^2
$$
denotes the Schwarz derivative. 
We have
\begin{equation}
   p(x)   \;=\;
 \cases{\sqrt{x} \left( 1 + x^2/5 + O(x^4) \right)\, & for $x\rightarrow 0$,\cr
           \noalign{\vskip 4pt}
           -\frac{1}{4} \log{ (1-x)} + \log{2}/2 +\pi/8 + O (x-1)\,
     & for $ x\rightarrow 1$,   \cr
         }
   \label{exact_fd}
\end{equation}
and so 
\begin{equation}
   \zeta (x)   \;=\;
 \cases{  x - 2 x^2/3 + O(x^3)\,   & for $x\rightarrow 0$,\cr
           \noalign{\vskip 4pt}
     1 - 2 \sqrt{1-x} e^{-\pi/4}\,       
     & for $ x\rightarrow 1$.   \cr
         }
   \label{exact_fd_1}
\end{equation}
Accordingly, the potential ${\cal P}(\zeta )$ becomes
\begin{equation}
  {\cal P}(\zeta )  \;=\;
 \cases{   (\nu^2-1)/4\zeta^2 + O(1)\,  & for $x\rightarrow 0$,\cr
           \noalign{\vskip 4pt}
         -1 / 4(1- \zeta )^2  + O(1)\,  
     & for $ x\rightarrow 1.$   \cr
         }
   \label{exact_fd_2}
\end{equation}
Introducing the ``regularized'' potential,
${\cal P}_R = {\cal P} - (\nu^2-1)/4\zeta^2 + 1/4 (1-\zeta)^2$,
we can  write equation (\ref{U_eq}) as
\begin{equation}
U'' = \left( - \frac{\lambda^2}{4 \zeta (1-\zeta)^2} +
 \frac{\nu^2-1}{4\zeta^2} - \frac{1}{4(1-\zeta)^2} + {\cal P}_R\right) U\,,
\label{U_R}
\end{equation}
where ${\cal P}_R$
is a regular function for $\zeta \in [0,1]$.
We shall look for solutions of the equation (\ref{U_R}) in the form
$U = U_0 (1+h(\zeta ))$, where $U_0$ obeys
\begin{equation}
4\zeta^2(1-\zeta)^2 U_0'' +
 \left( \lambda^2 \zeta +\zeta^2 - (\nu^2-1)(1-\zeta)^2\right) U_0 =0 \,,
\label{U_0}
\end{equation}
and for the error-control function $h(\zeta )$ one can derive the following integral equation
\begin{equation}
h(\zeta ) = \int\limits_{\zeta}^{1} {\cal P}_R (1+ h(\zeta )) U_o^2(\xi)
 \int\limits_{\xi}^{\zeta}\frac{dt}{U_0^2(t)} d\xi \,.
\label{error_control}
\end{equation}
The solution of (\ref{U_0}) is given by
\begin{equation}
U_0 (\zeta ) = (1-\zeta )^{\frac{1-i\lambda}{2}}(-\zeta )^{\frac{1+\nu}{2}}
\ofo \left( \frac{\nu -i\lambda}{ 2} +  \eta_+ (\nu),
\frac{\nu  -i\lambda}{ 2} + \eta_- (\nu);
 1- i\lambda;1-\zeta \right)\,,
\end{equation}
where
$\eta_{\pm}(\nu ) = \left( 1 \pm \sqrt{\nu^2-1} \right)/2$.
Returning to the $x$ variable, the solution normalized to $(1-x)^{-i\lambda /4}$ for $x\rightarrow 1$ is 
\begin{eqnarray}
\phi_0 (x) &=& C_{\lambda} x^{3/4}\cosh^{i\lambda}p(x) \tanh^{\nu+1/2} p(x)
\nonumber \\ &\,&
\ofo \left(  -i\lambda /2 + \nu /2 + \eta_+, -i\lambda /2 +\nu /2 +\eta_-;
 1- i\lambda;
 \mbox{sech}^2 p(x) \right)\,,
\end{eqnarray}
where $C_{\lambda} = \exp{(i\lambda \log{2}/2 - i\pi \lambda /8)}$.
For small $x$ (recall that $x=\rho_0^2/\rho^2$) one has
\begin{equation}
\phi_0 \rightarrow \frac{\rho^l}{\rho_0^l}
\frac{ C_{\lambda}\;\Gamma (\nu)\;\Gamma (1-i\lambda )}{
\Gamma \left( -i\lambda /2 + \nu /2 \eta_+ (\nu) \right)
\Gamma \left( -i\lambda /2 + \nu /2 + \eta_- (\nu) \right)}\,.
\end{equation}
This matches (for $\omega R \ll 1$) the outer region solution,
$\alpha J_{\nu}(\rho)/\rho^2 \sim \alpha \rho^l /2^{l+2}\Gamma (l+3)$, with
\begin{equation}
\alpha = \frac{ 2^{\nu} C_{\lambda}\Gamma (\nu +1 )
\Gamma (\nu)\Gamma (1-i\lambda )}{
\Gamma \left( -i\lambda /2 +\nu /2 + \eta_+ (\nu) \right)
\Gamma \left( -i\lambda /2 + \nu /2 +\eta_- (\nu) \right)}\,.
\end{equation}
The absorption probability 
(\ref{abs_prob})  can be written as
\begin{equation}
P = \frac{2\pi \lambda}{[\nu!(\nu -1)!]^2}
\frac{|\Gamma \left( \nu /2 + \eta_+ (\nu ) +  i\lambda /2 \right)
 \Gamma \left( \nu/2 + \eta_- (\nu ) +  i\lambda /2 \right)|^2}
{|\Gamma (1+  i\lambda )|^2} \left( \frac{\omega r_0}{2}\right)^{2\nu}
\label{absorption_hyper}
\end{equation}
where we used 
$|\Gamma (x + iy)| = |\Gamma (x - iy)|$.
The zero-temperature limit (corresponding to $\lambda
 \rightarrow \infty$) gives (see Appendix \ref{appendix_b})
\begin{equation}
P = P_0  \left( 1+ 4\nu (\nu^2-1)/3\lambda^2 +\cdots \right) \,,
\label{inc}
\end{equation}
where $P_0$ is given by (\ref{zero_temp_abs}).
We must not forget, however, that there are corrections to the result
(\ref{absorption_hyper}) coming from the function $h(\zeta )$ in (\ref{error_control}). Standard analysis of the Volterra equation  (\ref{error_control})
suggests that  $h(\zeta )$ is indeed a small correction (properties of ${\cal P}_R$ and $U_0$ ensuring convergence of corresponding integrals at the endpoints) of the order of $1/\lambda^2$. Therefore only the zeroth order term in
the expansion (\ref{inc}) can be trusted and thus there is no 
contradiction with our main result (\ref{d3_result}). 
The significance of (\ref{absorption_hyper}) is twofold: it shows that the absorption probability must contain contributions of the form $\exp{(-\lambda)}$,
and that it has a functional form expected in a field theory (compare
(\ref{p_exact})).
We can try to guess the exact expression for the absorption probability by
choosing the ansatz of the form\footnote{For the $D1-D5$ system
this method gives the exact thermal factor in 
(\ref{p_exact}) immediately.}
\begin{equation}
P = \frac{2\pi \lambda }{[\nu!(\nu -1)!]^2}
\frac{|\Gamma \left( x_1 +  i\lambda /2  \right)
 \Gamma \left( x_2 +  i\lambda /2  \right)|^2}
{|\Gamma (1+ i\lambda )|^2} \left( \frac{\omega r_0}{2}\right)^{2\nu}\,
\label{ans}
\end{equation}
Comparison with the asymptotic expansion  (\ref{d3_result})
  to the order $1/\lambda^2$ 
gives 
$$
x_{1,2} = \left( \nu + 1 \pm \sqrt{(1-\nu^2)/3}\right)/2\,.
$$
However, the coefficient in front of $1/\lambda^4$ term turns out to be
$32\nu (\nu^2-1)(4\nu^2-1)/45$ which does not match  (\ref{d3_result}).
This suggests that the naive thermal ansatz (\ref{ans}) is incorrect.

\subsection{High-temperature expansion}\label{D3_high_T}
Two independent solutions of (\ref{x_eq}) are
\begin{equation}
\phi_{\pm} (x) = (1-x )^{\sigma_{\pm}} F(x)\,,
\end{equation}
where $\sigma_{\pm}=\pm i \lambda /4$
 are the roots of the indicial equation and $F(x)$ obeys
\begin{eqnarray}
\frac{d^2 F}{d x^2} & - &
 \frac{x^2(1+2\sigma ) + 
2\sigma x +1}{x (1-x^2)}\frac{d F}{d x} 
- \frac{l(l+4)}{4 x^2 (1-x^2)} F \nonumber \\
 &+& \frac{x (1+x )\sigma (\sigma -1) +
\sigma (1+x^2)}{x (1-x)^2(1+x )} F +
\frac{\lambda^2}{4 x (1-x^2)^2} F=0\,.
\label{xi_mod}
\end{eqnarray}
The solution corresponding to the incoming wave at the horizon is
\begin{equation}
\phi (r)= A (1- r_0^2/r^2 )^{-i\lambda /4} F(r)\,,
\end{equation}
where $A$ is an arbitrary constant and $F(r)$ is
 supposed to be regular at $r=r_0$ ($x=1$).
For $\lambda \ll 1$ the approximation scheme can be constructed applying 
 a standard
perturbation theory to the equation (\ref{xi_mod}).
In the zeroth-order approximation (\ref{xi_mod})
reads
\begin{equation}
x^2(1-x^2)F'' - x(1+x^2)F' - l(l+4)/4 \, F =0\,,
\end{equation}
and has solutions
\begin{eqnarray}
F_1 (x) & = & x^{2+l/2} \ofo \left( 1+l/4, 1+l/4; 2+l/2;x^2\right)\,,
\nonumber \\
F_2 (x) & = & x^{-l/2} \ofo \left( -l/4, -l/4; -l/2; x^2\right)\,.
\end{eqnarray}
We have to distinguish between even and odd $l$.
\subsubsection*{Odd $l$:}
The solution regular at $x=1$ (i.e. the one without 
logarithmic singularities) is given by
$F(x) = F_1(x) + C_l F_2 (x)$, where
$$
C_l = - \frac{\Gamma (2+\frac{l}{2} )
\Gamma^2 ( -\frac{l}{4})}{\Gamma ( -\frac{l}{2})
 \Gamma^2 ( 1+\frac{l}{4})}
$$
Matching this with the outer region's solution gives
$\alpha = 2^{l+2} (l+2)! C_l/\rho_0^l$,
and the absorption probability
\begin{equation}
P =  P_0(l) \left( \frac{2}{\lambda}\right)^{2l+3}
\frac{(l+1)!^2}{\pi |C_l|^2} = 
P_0(l) \left( \frac{2\pi T}{\omega}\right)^{2l+3}
\frac{(l+1)!^2}{\pi |C_l|^2}\,,
\label{high_T_odd}
\end{equation}
where $P_0(l)$ is given by (\ref{zero_temp_abs}).
\subsubsection*{Even $l$:}
In this case the solution regular at $x=1$ is $F_2(x)$.
For $l=4n$, $n=0,1,2,...$, it is given by
$$
F_2(x) = \frac{n!^2}{(2n)!} P_n \left( \frac{2}{x^2}-1\right)\,,
$$
where $P_n(z)$ are the Legendre polynomials, and we have
$F_2(x)\rightarrow x^{-l/2}$ for $x\rightarrow 0$.
For  $l=4n+2$, $n=0,1,2,...$ the solution can be written in terms of the 
Meijer function \cite{prudnikov}:
$$
F_2(x) = G^{2 0}_{2 0} \left( x^2 \left| \begin{array}{cccc}
\;\; \;\;\;\cdot \; \;\;\;\hspace{0.8cm}\;\; \cdot \;\hspace{1.3cm} \; 1\; 1 \\
  -n-1/2 \;\; n+3/2 \;\hspace{0.3cm}\;\;  \cdot \; \cdot
 \end{array}\right)\right.\,.
$$
We have $F_2(x) \rightarrow x^{-l/2}/\Gamma^2 (1+l/4)$ for 
 $x\rightarrow 0$. 
 Accordingly, the absorption probability is
\begin{equation}
P = P_0(l) \left( \frac{2}{\lambda}\right)^{2l+3}
\frac{(l+1)!^2}{\pi} = 
P_0(l) \left( \frac{2\pi T}{\omega}\right)^{2l+3}
\frac{(l+1)!^2}{\pi}\,,
\label{high_T_even_0}
\end{equation}
for $l=0,4,8,...$ and 
\begin{equation}
P = P_0(l) \left( \frac{2}{\lambda}\right)^{2l+3}
\frac{(l+1)!^2\Gamma^4(1+l/4)}{\pi} = 
P_0(l) \left( \frac{2\pi T}{\omega}\right)^{2l+3}
\frac{(l+1)!^2\Gamma^4 (1+l/4)}{\pi}\,,
\label{high_T_even_2}
\end{equation}
for $l=2,6,10...$.
Thus for  the $s$-wave the absorption cross section 
$\sigma_0 = 32\pi^2 P(0)/\omega^5$ can be written 
as $\sigma_0 = A(r_0, R) f(\lambda )$, where $A(r_0,R) = \pi^3r_0^3R^2$
is the horizon area of the three-brane\footnote{This 
is reminiscent of the universal result
 for the black holes \cite{Das:1997we}.}
 in the near-extremal limit,
 and $f(\lambda)$ 
interpolates between 1 at $\lambda =0$ and $\pi \lambda^3/8$ at
 $\lambda =\infty$.

\noindent
Corrections to the lowest-order results  
(\ref{high_T_odd}), (\ref{high_T_even_0}), (\ref{high_T_even_2})
can be obtained by applying regular perturbation theory to the equation
(\ref{xi_mod}). For example, for the $s$-wave we have
\begin{equation}
F(x) = 1 - \frac{i\lambda}{4}\log{\left( \frac{1+x}{2}\right)} + 
\frac{\lambda^2}{32}
\left[ \left( \log{\left( \frac{1+x}{2}\right)} +8\right) 
 \log{\left( \frac{1+x}{2}\right)} - 4
\, \mbox{Li}_2\left(\frac{1-x}{2}\right)\right] + O(\lambda^3)\,.
\end{equation}
Correspondingly, the absorption cross section becomes
\begin{equation}
\sigma_0 = \pi^6 R^8 T^3 \left[ 1 + \frac{
\left(\pi^2+12(2-\log{2})\log{2}\right)\omega^2}{48\pi^2 T^2}
 + \cdots \right]\,.
\label{high_T_corr}
\end{equation}
The frequency-dependent term in (\ref{high_T_corr}) 
can be interpreted as a correction to a hydrodynamic
description of the strongly coupled Yang-Mills plasma \cite{Son}.

\section{Absorption by the  $D1/D5$-branes}\label{D1D5}
\noindent
The bound state of $D1$ and $D5$ branes has been intensively
 studied as a string theory model for a five-dimensional black hole, and
later in the context of AdS/CFT correspondence 
(discussion and references can be found in  \cite{Aharony:2000ti}). 
The non-extremal absorption probability contains a thermal factor 
which can be calculated exactly using 
a (super)gravity description in the ``dilute gas'' regime.
Here we would like to calculate this temperature-dependent coefficient
by using Langer-Olver's approximation. We shall see that power-like 
terms at any given order of
 the low-temperature expansion are indeed correctly reproduced,
but the  corrections 
exponentially suppressed at low temperatures are 
not captured.
We start with the ``near region'' equation (3.7)
 of \cite{Maldacena:1997ih},
 where we shall
put $T_L  = T_R = T_H = T$ for simplicity and use  
notations $x=r_0^2/r^2$, $\lambda = \omega/2\pi T$: 
\begin{equation}
\frac{d^2 \phi}{d x^2}  - \frac{1}{(1-x)}
 \frac{d \phi }{d x} - \frac{l(l+2)}{4 x^2 (1-x)} \phi +
\frac{\lambda^2}{4 x (1-x)^2} \phi =0\,,
\label{d1_d5_eq}
\end{equation}
Introducing a new function $u$ by 
$u=\sqrt{x-1}\, \phi$ one can write (\ref{d1_d5_eq}) in the form
\begin{equation}
u'' = \left( \lambda^2 F + G \right) u\,,
\label{u_d1_d5}
\end{equation}
where 
$F = -1/4 x(1-x)^2$, $G= l(l+2)/4x^2(1-x) - 1/4(1-x)^2$.
Equation (\ref{d1_d5_eq}) has an exact solution given (up to an arbitrary constant) by
\begin{equation}
\phi = x^{-l/2}(1-x)^{-i\lambda /2}\, \ofo \Biggl( -(l+i\lambda)/2, 
-(l + i\lambda )/2; 1-i\lambda ;1-x \Biggr)
\label{exact_d1_d5}
\end{equation}
which leads to the absorption probability \cite{Maldacena:1997ih}
\begin{equation}
P(l) = \frac{2\pi \lambda}{l!^2(l+1)!^2}
 \left( \frac{\omega r_0}{2}\right)^{2l+2} \left| \frac{\Gamma^2
 \left( \frac{l+2}{2} - \frac{i\lambda }{2}\right)}
{\Gamma \left( 1-i\lambda\right)} \right|^2\,.
\label{p_exact}
\end{equation}
The plane wave absorption cross-section is then obtained by multiplying 
(\ref{p_exact}) by $4\pi/\omega^3$.
In the low temperature limit ($\lambda \rightarrow \infty$) we have
\begin{eqnarray}
P(l)/P_0(l) & = & 1 + \frac{\nu (\nu^2-1)}{3\lambda^2} 
+\frac{\nu (\nu^2-1)(5\nu^3 - 9\nu^2 - 5\nu +21)}{90\lambda^4}\nonumber \\
& + & \frac{\nu (\nu^2 -1)(\nu -2)(\nu -3)(35\nu^4 -14\nu^3 -
 80 \nu^2 +314 \nu +465)}{5670\lambda^6}\nonumber \\
 & + & O\left( 1/\lambda^6\right) +  
O\left( e^{-\lambda}\right)\,,
\label{p_00}
\end{eqnarray}
where $\nu = l+1$, the absorption probability at zero temperature is given by
\begin{equation}
P_0 (l)=
\frac{4\pi^2}{l!^2(l+1)!^2} 
\left( \frac{\lambda \omega r_0}{4}\right)^{2l+2}
 =
\frac{\pi^2 \omega^4 r_1^2 r_5^2}{4}  \frac{(\omega^2 r_1 r_5)^{2l}}
{2^{4l}l!^2(l+1)!^2}\,,
\end{equation}
and we have used $T=r_0/2\pi r_1 r_5$.
Notice that in addition to the power-like corrections coming from
gamma-functions in the numerator of (\ref{p_exact}), 
expression (\ref{p_00}) also contains corrections of the form
 $\exp{(-\lambda)}$ produced by $|\Gamma (1-i\lambda )|$ in 
the denominator
(see Appendix \ref{appendix_b}).
 In fact, for $l=0$ there are no power-like corrections at any order:
\begin{equation}
P(0) = \left( \frac{\pi\lambda \omega r_0}{2}
\right)^2 \coth{ \frac{\pi\lambda}{2}} = 
P_0 (0)\left( 1 + 2 e^{-\pi \lambda} + \cdots \right)\,.
\end{equation}

\subsection{Low-temperature asymptotic expansion}\label{D1D5_low_T}

\noindent
Using Langer-Olver's method, 
we would like to construct an approximate solution of
 (\ref{d1_d5_eq}) in the form of asymptotic series in powers of $1/\lambda^2$, 
and compare it to the large $\lambda$ expansion of the exact solution.
We shall use the notations of Section \ref{D3} throughout, with the 
exception of $\lambda = \omega/2\pi T$ and $\nu = l+1$.
Applying the conformal transformation 
\begin{equation}
\zeta = - \mbox{arctanh}^2 \sqrt{x} \,,
\end{equation}
\begin{equation}
W = \frac{\mbox{arctanh}^{1/2} \sqrt{x}}{x^{1/4}\sqrt{x-1}} u\,,
\label{W_u_d1_d5}
\end{equation}
we can write (\ref{u_d1_d5}) as
\begin{equation}
\frac{d^2 W}{d\zeta^2} = \left\{ \frac{\lambda^2}{4\zeta} + 
\frac{\nu^2-1}{4\zeta^2} + \frac{\psi (\zeta )}{\zeta}\right\} W\,,
\label{W_d1_d5}
\end{equation}
where the function
\begin{equation}
\psi (x) = \frac{1-4\nu^2}{16 \zeta (x)} + \frac{(1-x)(1-4 \nu^2 - x)}{16 x}
\end{equation}
is analytic at $x=0$. The
 formal series solution of (\ref{W_d1_d5})
 representing the incoming wave is given by
\begin{eqnarray}
W (x) &=& \frac{i^{l+1}}{N_1 (\lambda ,l)} \sqrt{\frac{\pi\lambda}{2}}
  \mbox{arctanh} 
\sqrt{x}\; \Biggl\{ H^{(1)}_{l+1} (\lambda\; \mbox{arctanh}\sqrt{x})
 \sum\limits_{n=0}^{\infty} A_n (x) \lambda^{-2n} 
\nonumber \\ &-& \frac{\lambda}{\mbox{arctanh}\sqrt{x}} H^{(1)}_{l+2}
 (\lambda \;
\mbox{arctanh}\sqrt{x}) \sum\limits_{n=0}^{\infty} B_n (x) 
\lambda^{-2n}\Biggr\}\,,
\label{W_solution}
\end{eqnarray}
where the 
normalization constant $N_1(\lambda ,l)$,
\begin{equation}
N_1(\lambda ,l) =  e^{i\lambda \log{2} - i\pi /4}
 \sum\limits_{n=0}^{\infty}\left( A_n (1) +\frac{i}{\lambda}
 \lim_{x\rightarrow 1}\mbox{arctanh}\sqrt{x} B_n(x)\lambda^{-2n}\right)\,,
\end{equation}
is chosen in such a way that
\begin{equation}
\bar{\phi} =
 x^{1/4}\mbox{arctanh}^{-1/2}\sqrt{x}\; W \rightarrow (1-x)^{-i\lambda /2}
\end{equation}
for $x\rightarrow 1$, and thus $\bar{\phi}$ 
can be identified with the exact solution
(\ref{exact_d1_d5}).
Since for $x\rightarrow 0$
\begin{equation}
\bar{\phi} \rightarrow
 \frac{N_0\, 2^l\, i^l\, l!}{N_1 \lambda^{l+1}}\; x^{-l/2}\,,
\end{equation}
where
\begin{equation}
N_0 (\lambda ,l) = \sum\limits_{n=0}^{\infty}\lambda^{-2n}
\left( A_n(0) - 2\nu B_n(0)\lambda^{-2}\right)\,,
\end{equation}
one can match with the outer region's solution $\alpha J_{l+1}(\rho)/\rho$
in the overlapping region $\rho \ll 1$,
and obtain the absorption probability
\begin{equation}
P(l) = 
\frac{4\pi^2}{l!^2(l+1)!^2} 
\left( \frac{\lambda \omega r_0}{4}\right)^{2l+2}
 \left| \frac{N_1}{N_0}\right|^2 =
 P_0(l)  \left| \frac{N_1}{N_0}\right|^2\,.
\label{olver_abs}
\end{equation}
The recursion relations for the coefficients 
(\ref{bar_a}) - (\ref{bar_b}) are now
\begin{equation}
\bar{A}_{k+1}(x) = 
i\sqrt{x}(x-1)\partial_x 
\bar{A}_k + i\int \frac{4\nu^2 - 1 + x}{16x\sqrt{x}} \bar{A}_k dx\,.
\end{equation}
One finds
\begin{equation}
B_0(x) = \frac{1-4\nu^2}{8\mbox{arctanh}^2\sqrt{x}} -
\frac{\sqrt{x}}{8\mbox{arctanh}\sqrt{x}}
 \left(1+ \frac{1-4\nu^2}{x}\right)\,,
\end{equation}
so that $B_0(0) = l(l+2)/6$, and
$\lim_{x\rightarrow 1} \mbox{arctanh}\sqrt{x} B_0(x) =
 \nu^2/2 -1/4$.
Limiting values of a few higher-order coefficients can be found in Appendix \ref{appendix_a}. 
Substitution into (\ref{olver_abs}) gives precisely (\ref{p_00}) as expected.
Due to the simplicity of the function $\psi(x)$, 
the calculation of coefficients $A_n$, $B_n$  
is straightforward,
and therefore our result can in principle be
 extended to all orders in $1/\lambda^2$. 
However, the asymptotic expansion (\ref{W_solution}) cannot capture
$\exp{(-\lambda)}$ correction terms which in the
 $s$-wave
case would lead to an erroneous
 conclusion that $P(0) = P_0(0)$ for non-zero temperature.

\begin{figure}[p]
\begin{center}
\epsffile{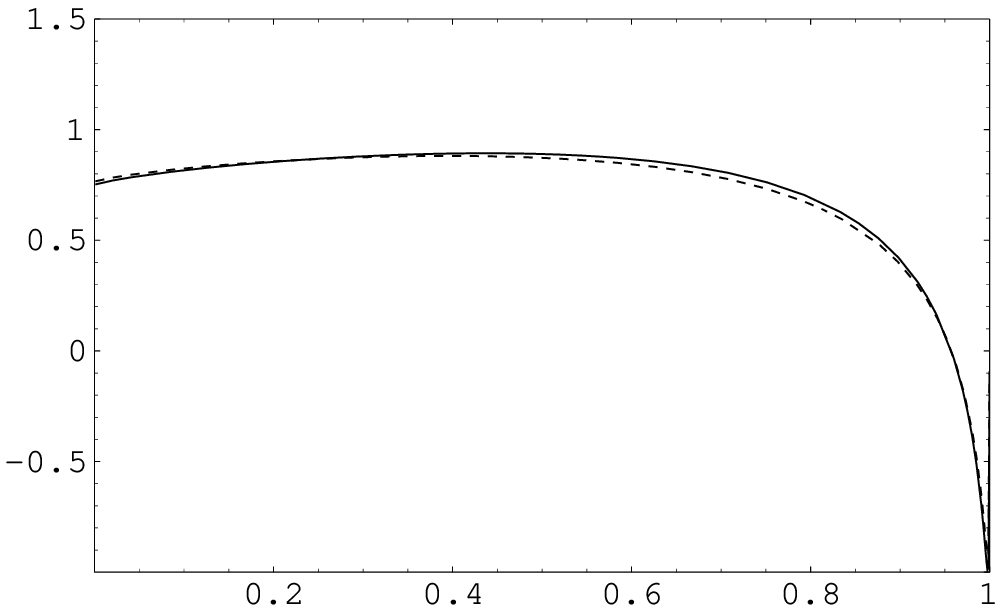}
\end{center}
\caption{$D1/D5$ system: exact solution Re $\phi (x)$ (solid line) and
a partial sum (with all $A_s$, $B_s$ equal to zero for $s>0$) 
of the asymptotic series for $\lambda = 1$ (dashed line).
}
\label{solutions_d1_d5}
\end{figure}

\begin{figure}[p]
\begin{center}
\epsffile{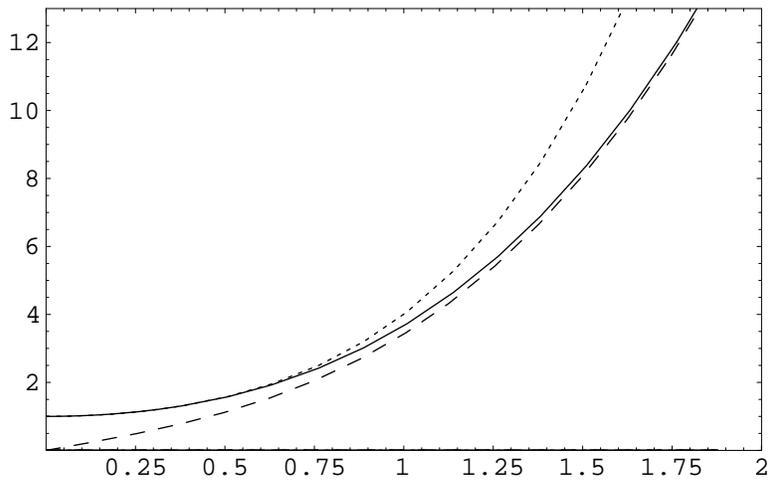}
\end{center}
\caption{$D1/D5$ system: ratio $P(l)/P_0(l)$ versus $1/\lambda$ 
for $l=1$: exact (solid line),
high-temperature approximation (\ref{d1_d5_T_1}) (dashed line), 
sum of the first three terms of 
the low-temperature asymptotic 
expansion (\ref{p_00}) (dotted line).
}
\label{d1_d5_compa}
\end{figure}

In Figure~(\ref{solutions_d1_d5}) we plot the exact and the approximate
solutions of  (\ref{d1_d5_eq}).
Figure~(\ref{d1_d5_compa}) shows the ratio  $P(l)/P_0(l)$
 (exact and approximate results) as a function
of $1/\lambda$. It is clear that the Langer-Olver approximation 
which is supposed to work well for large $\lambda$ is in fact 
quite good even for $\lambda \sim 1$. 
We remark that the 
Pade-improvement of the low-temperature asymptotic series  (\ref{p_00})
does not seem to be particularly helpful in this case.

\subsection{High-temperature expansion}\label{D1D5_high_T}

\noindent
The leading term in the high-temperature expansion
 ($\lambda \ll 1$) of (\ref{p_exact}) is
\begin{equation}
P = P_0(l) \left( \frac{2}{\lambda}\right)^{2l+1}
\frac{\Gamma^4(1+l/2)}{\pi} = 
P_0(l) \left( \frac{4\pi T_H}{\omega}\right)^{2l+1}
\frac{\Gamma^4 (1+l/2)}{\pi}\,.
\end{equation}
For the $s$-wave the absorption cross section is given by
$\sigma_0 = 2\pi^2 r_0 r_1 r_5 f(\lambda)$, where $f(\lambda)=(\pi \lambda /2)
\coth{(\pi \lambda /2)}$ interpolates 
between $1$ at $\lambda = 0$ and $\pi\lambda /2$ at $\lambda =\infty$.
Corrections can be readily obtained by solving equation 
(\ref{d1_d5_eq}) perturbatively in $\lambda \ll 1$.
For the $s$-wave we get
\begin{equation}
\phi (x) = (1-x)^{-i\lambda /2}\left( 1 - \frac{\lambda^2}{4}
 \mbox{Li}_2 (1-x) +\cdots \right)\,.
\end{equation}
Correspondingly, we have for $l=0$:
\begin{equation}
P(0)/P_0(0) = 2/\pi\lambda + \pi\lambda/6 + O(\lambda^2)\,.
\label{d1_d5_T_0}
\end{equation}
For $l=1$ the result is
\begin{equation}
P(1)/P_0(1) = \frac{\pi}{2\lambda^3} + 
\frac{\pi}{\lambda}\left( 1 - \frac{\pi^2}{24}\right) + O(\lambda )\,.
\label{d1_d5_T_1}
\end{equation}
Of course, formulae (\ref{d1_d5_T_0}), (\ref{d1_d5_T_1}) coincide with the 
$\lambda \ll 1$ Taylor expansion of the exact result (\ref{p_exact}).

\section{Field theory calculation}\label{field_theory}

\noindent
In field theory, the absorption cross section at finite temperature 
can be calculated using the method of thermal
 Green's functions (see e.g. \cite{abrikosov}).
 The absorption cross section is determined by a
 discontinuity of the two-point function of the operator ${\cal O}_l$ 
to which the $l$th partial wave of the dilaton field couples, 
and is given by the zero-momentum limit of the spectral density,
\begin{equation}
\sigma_l^T = \lim_{k\rightarrow 0} \rho_l^T (\omega , k)\,,
\label{abs_thermal}
\end{equation}
where $\rho_l^T (\omega , k)$
at finite temperature in the imaginary time formalism is proportional to 
the discontinuity of the (analytically continued) Euclidean thermal 
two point function $\Pi (r, \tau )$,
\begin{equation}
\rho_l^T (\omega , k) = \frac{\Omega}{{\cal F}}\int d^3x
 dt e^{-i{\bf k}{\bf x} +i \omega t}
 \left[ \bar{\Pi}(r, t-i\epsilon ) -  
  \bar{\Pi}(r, t+i\epsilon ) \right]\,.  
\label{spectral_density}
\end{equation}
Here the function $\bar{\Pi}(r, t)$ denotes the  analytic continuation of
 $\Pi (r, \tau )$ to the real time $t=-i\tau$, ${\cal F}$ denotes 
the flux, and $\Omega$ is a coefficient 
 which depends on the parameters of the system under consideration.

\subsection{D1/D5 system}
\noindent 
The absorption cross section is given by 
(\ref{abs_thermal}), (\ref{spectral_density}), where
$\bar{\Pi}(r,t)$ is the (analytically continued) two-dimensional thermal
 Green's function,
\begin{equation}
\bar{\Pi}(r,t) = \frac{{\cal C_O}}{i^{2h_L + 2h_R}}
\left( \frac{\pi T_L}{\sinh{\pi T_L(t+x)}}\right)^{2h_L}
 \left( \frac{\pi T_R}{\sinh{\pi T_R(t-x)}}\right)^{2h_R}\,.
\end{equation}
The integral in 
(\ref{spectral_density}) factorizes nicely by introducing $x_{\pm}=t\pm x$,
and can be computed exactly \cite{Gubser:1997cm}.
This factorization is a property of a two-dimensional theory and does
not hold in higher dimension. To illustrate the approach we shall use 
in the $d=4$ case we compute the $D1/D5$ absorption cross section 
$\sigma_0^T$ by integrating over $t$ and $x$ directly without factorization.
One can also show that the low-temperature asymptotic series (\ref{p_00}) 
can be obtained by expanding $\bar{\Pi}(r,t)$ in powers of $T$.
\noindent
The spectral density for the $s$-wave is given by 
(\ref{spectral_density}) with $\Omega = 2\pi r_1^2 r_5^2$,  ${\cal F}=\omega$ and
\begin{equation}
\bar{\Pi}(r,t) = \frac{\pi^4 T_L^2 T_R^2}{\sinh^2 \pi T_L (t+x) 
\sinh^2 \pi T_R (t-x)}\,.
\end{equation}
Integrating over time we get
\begin{eqnarray}
\rho_0^T (\omega , k) &=& \frac{4\pi^4 r_1^2 r_5^2}{\omega}
\int dx e^{- i k x}\Biggl[
 \frac{  e^{-i\omega x}  T_R^2
\left(2\pi i T_R \coth{(2\pi T_R x}) - \omega\right)}
{\sinh^2 2\pi T_R x} \nonumber \\
&-&
\frac{ e^{i\omega x} T_L^2\left( 2\pi i T_L \coth{(2\pi T_L x)} + 
\omega\right)}
{\sinh^2 2\pi T_L x}\Biggr]\,.
\end{eqnarray}
Evaluating residues, we obtain the spectral density
\begin{equation}
\rho_0^T (\omega , k) = \frac{\pi^2(\omega^2-k^2)}{4}
\left[ \coth{\frac{(\omega + k)}{4T_R}} + \coth{\frac{(\omega - k)}
{4T_L}}\right]\,.
\label{rho_s}
\end{equation}
The absorption cross section is then obtained by taking 
the limit $k\rightarrow 0$ in (\ref{rho_s}),
\begin{equation}
\sigma = \sigma_0 f(\omega , T_L, T_R) =
 \pi^3 r_1^2 r_5^2\omega
 \frac{e^{\frac{\omega}{T_H}} -1}{
\left(e^{\frac{\omega}{2T_L}} -1\right)
\left(  e^{\frac{\omega}{2T_R}} -1\right)}\,,
\label{abs_d1_d5_s}
\end{equation}
where  $\sigma_0$ is the cross-section at zero temperature. 
Similar calculations can be done for higher 
partial waves. For all of them one observes a perfect agreement with 
the results obtained from gravity 
\cite{Das:1996jy}, \cite{Das:1996wn}, \cite{Maldacena:1997ih}, 
\cite{Gubser:1997cm}.

\subsection{Non-extremal three-branes}
\noindent
To the
 leading order in 't Hooft coupling, the terms contributing
 to $\langle {\cal O}_l(x){\cal O}_l(0)\rangle$ are a bosonic term
$\Pi_l^{\mbox{\tiny{bos}}}$ (for all $l\geq 0$), a two-fermion term
 $\Pi_l^{2\Theta}$ (for $l\geq 1$), and a four-fermion term
  $\Pi_l^{4\Theta}$ (for $l\geq 2$) identified in
\cite{Klebanov:1999xv}:
\begin{equation}
\Pi_{l,\mbox{\tiny{bos}}}^{T=0}(x)
 = \frac{\kappa^l \omega^{2l}N^{l+2}(l+2)(l+3)}
{3\pi^{l/2} 2^{l+1}l!(l+1)!}\Delta^l(x)\left( \partial_a\partial_b \Delta(x)
\partial_a\partial_b \Delta(x) +\frac{1}{2}\partial^2\Delta (x)
 \partial^2\Delta (x)\right)
\label{bos_contribution}
\end{equation}
where $\Delta (x) = 1/4\pi^2|x|^2$ is the free scalar propagator.
The term proportional to $\partial^2\Delta (x)$ does not contribute to the
discontinuity and will be omitted.
At finite temperature the bosonic contribution to the two-point function
will be given by (\ref{bos_contribution}) with  $\Delta (x)$ replaced by
its thermal cousin. 
Borrowing the expression for the thermal 
propagator in  position space from \cite{Vshivtsev:1991aj},
\begin{equation}
{\cal G}_B (r,\tau ) = \frac{T}{4\pi r}
 \frac{\sinh{ 2\pi r T}}{(\cosh{ 2\pi r T}-\cos{ 2\pi \tau T})}\,,
\end{equation}
where $r=|\vec{x}|$,
we obtain
\begin{equation}
\Pi_{l,\mbox{\tiny{bos}}}^{T}(r, \tau )
 = \frac{\kappa^l \omega^{2l}N^{l+2}(l+2)(l+3)}
{3\pi^{l/2} 2^{l+1}l!(l+1)!}{\cal M}_l (r,\tau )\,,
\label{the_bos}
\end{equation}
where
\begin{equation}
{\cal M}_l (r,\tau ) = {\cal G}_B^l\left[ (\partial^2_{\tau\tau}
 {\cal G}_B)^2 + 
(\partial^2_{rr} {\cal G}_B)^2 + 2(\partial^2_{r\tau} {\cal G}_B)^2 + 
2(\partial_{r} {\cal G}_B)^2/r^2\right]\,. 
\end{equation}
For the $s$-wave we obtain (see Appendix \ref{appendix_d} for details)
\begin{equation}
\rho_0^{\mbox{\tiny{bos}}}
(\omega , k) = \frac{\pi k (\omega^2-k^2)^2}{32}\left[
1 +\frac{2 T}{k}\log \left( \frac{ 1- e^{-\frac{\omega + k}{2T}} }
{ 1- e^{-\frac{\omega + k}{2T}}} \right)\right]\,.
\end{equation}
Taking the limit $k\rightarrow 0$ we get 
\begin{equation}
\sigma_0^T = 
\frac{\kappa^2 N^2 \omega^3}{32 \pi}\coth{\frac{\omega}{4T}} 
= \sigma_0^{T=0}\coth{\frac{\omega}{4T}}\,,
\end{equation}
where $\kappa^2 N^2 = 4\pi^5 R^8$.

\noindent
For partial waves with $l>0$ in addition to (\ref{spectral_density})
one has to take into account contributions of
the two- and four-fermion operators. 
The contribution of the two-fermion operator is
\begin{equation}
\Pi_{l, 2\Theta}^{T}(r, \tau )
 =  \frac{\kappa^l \omega^{2l}N^{l+2}(l+2)(l+3)}
{3\pi^{l/2} 2^{l}(l-1)!(l+1)!}{\cal M}_{l, 2\Theta} (r,\tau )\,,
\end{equation}
where 
\begin{equation}
{\cal M}_ {l, 2\Theta}(r,\tau ) = {\cal G}_B^{l-1}
\left[ \partial^2_{rr}
 {\cal G}_B (\partial_{r} {\cal G}_F)^2 
+
 \partial^2_{\tau \tau}
 {\cal G}_B (\partial_{\tau} {\cal G}_F)^2 
+ 2  \partial^2_{r\tau}
 {\cal G}_F \partial_{r} {\cal G}_F  \partial_{\tau} {\cal G}_F 
\right]\,, 
\end{equation}
and the free fermionic propagator at finite temperature is given by
\begin{equation}
{\cal G}_F (r,\tau ) = \frac{T}{2\pi r} \frac{\sinh{ \pi r T}\cos{\pi T\tau}
}{(\cosh{ 2\pi r T}-\cos{ 2\pi \tau T})}\,.
\end{equation}
Here we only consider the 
$l=1$ case. 
For general $l$
 the calculation is similar but more tedious since the four-fermion operators
also contribute.
Expanding ${\cal M}_ {1, 2\Theta}(r,\tau )$ in powers of $T$ and evaluating
 the discontinuity we get 
\begin{equation}
\sigma_{1,\mbox{\tiny field th}}^T
 = \frac{\kappa^3 N^3 \omega^7}{6144\pi^{7/2}}\left[
1+ 8\left(\frac{\pi T}{\omega}\right)^2 + 48
\left(\frac{\pi T}{\omega}\right)^4 + 
32\left(\frac{\pi T}{\omega}\right)^6 + O\left( e^{-\omega /T} \right)\right]\,.
 \end{equation} 
Comparing this result with the one obtained from gravity,
\begin{equation}
\sigma_{1,\mbox{\tiny grav}}^T
 = \frac{\kappa^3 N^3 \omega^7}{6144\pi^{7/2}}\left[
1 + 48
\left(\frac{\pi T}{\omega}\right)^4  + O\left( e^{-\omega /T} \right)\right]\,,
 \end{equation} 
we observe that $\sigma_{1,\, \mbox{\tiny field th}}^T$ contains
 terms proportional to $T^2$ and $T^6$ which are absent in
 $\sigma_{1,\, \mbox{\tiny grav}}^T$.

\section{Black $p$-branes and non-extremal $M$-branes}\label{M_branes}
    
\noindent
In this Section we demonstrate that the $M$- and black $p$-brane backgrounds
can be treated by the approximation methods used in Section 2
provided the correct identification of the expansion parameter is made.
\subsection{Non-extremal $M$-branes}
\noindent
Non-extremal $M$-brane solutions were found in \cite{Gueven:1992hh}.
In the notations of \cite{Cvetic:1996gq} the $M2$-brane metric is given by
\begin{equation}
ds^2_{11} = H^{-2/3}(r) \left( -f(r)dt^2 + dx_1^2 +dx_2^2\right)
 +H^{1/3}(r)\left( \frac{dr^2}{f(r)}+r^2 d\Omega_{7}^2\right)\,,
\label{M2_metric} 
\end{equation}
where $H(r) = 1 + R^6/r^{6}$, 
$f(r) =  1 - r_0^{6}/r^{6}$.
The radial part of the 
Laplace equation for a minimally coupled massless scalar
propagating in the background (\ref{M2_metric}) becomes
\begin{equation}
\frac{d^2 f}{d x^2}  - \frac{(5 x^3+1)}{2 x (1-x^3)}
 \frac{d f}{d x} - \frac{l(l+6)}{4 x^2 (1-x^3)} f +
\frac{\lambda^2}{(1-x^3)^2}
\left( 1 + \frac{\rho_0^6}{(\omega R)^6 x^3}\right) f =0\,,
\label{xx_eq_M2}
\end{equation}
where  $x= r_0^2/r^2$, 
$\lambda^2 = (\omega R)^6/4\rho_0^4$,  $\rho = \omega r$.
In the ``inner'' region $\rho \ll 1$ equation (\ref{xx_eq_M2}) 
is
\begin{equation}
w''(x)  =  \left(  - \frac{\lambda^2}{(1-x^3)^2} + 
\frac{l(l+6)}{4 x^2 (1-x^3)}
+\frac{5- 46 x^3 + 5 x^6}{16 x^2 (1-x^3)^2}\right) w(x)\,
\label{dirk_M2}
\end{equation}
where $w(x) = f(x) x^{-1/4}\sqrt{x^3-1}$.
Asymptotics of (\ref{dirk_M2}) can be constructed using Langer-Olver's
method. Here, however, the parameter of the expansion should be redefined
according to 
$\lambda^2 \rightarrow \Lambda^2 = \lambda^2 -3l(l+6)/4$ to ensure 
the convergence of the error-control function\footnote{This redefinition is
similar to the one performed in quantum mechanics where in
 quasi-classical approach to the central-force potentials a
substitution $l(l+1)\rightarrow (l+1/2)^2$ is made ``by hand'' 
to achieve the correct asymptotic behavior of the solution.}
\cite{Olver_book}.


\noindent
In the case of $M5$-branes the metric is given by
\begin{equation}
ds^2_{11} = \frac{-f(r)dt^2 + dx_1^2 +dx_2^2+dx_3^2 +dx_4^2+dx_5^2}
{H^{1/3}(r)}
 +H^{2/3}(r)\left( \frac{dr^2}{f(r)}+r^2 d\Omega_{4}^2\right)\,,
\label{M5_metric} 
\end{equation}
where $H(r) = 1 + R^3/r^{3}$, $f(r) =  1 - r_0^{3}/r^{3}$.
An analogue of the equation (\ref{xx_eq_M2}) is
\begin{equation}
\frac{d^2 f}{d x^2}  - \frac{(x^3-2)}{x (1-x^3)}
 \frac{d f}{d x} - \frac{l(l+3)}{ x^2 (1-x^3)} f +
\frac{\lambda^2}{x (1-x^3)^2}
\left( 1 + \frac{\rho_0^3}{(\omega R)^3x^3}\right) f =0\,,
\label{xx_eq_M5}
\end{equation}
where $x=r_0/r$, $\lambda^2 = (\omega R)^3/\rho_0$. In 
 the ``inner'' region,
i.e. for $\rho \ll 1$
equation (\ref{xx_eq_M5}) becomes
\begin{equation}
w''(x)  =  \left(  - \frac{\lambda^2}{x(1-x^3)^2} + \frac{l(l+3)}{x^2(1-x^3)}
+\frac{x(8-x^3)}{4(1-x^3)^2}\right) w(x)\,
\end{equation}
where $w(x) = f(x) x/(x^3-1)^{1/6}$. 
As before, an asymptotic expansion with the 
convergent coefficients can be built, 
provided the redefinition 
$\lambda^2 \rightarrow \Lambda^2 = \lambda^2 -4-3l(l+3)$
is made.

\subsection{Black $p$-branes:}
The metric is given by
\begin{equation}
ds^2_{10} = \frac{-f(r)dt^2 + d\vec{x}\cdot d\vec{x}}{\sqrt{H_p(r)}}
 +\sqrt{H_p(r)}\left( \frac{dr^2}{f(r)}+r^2 d\Omega_{8-p}^2\right)\,,
\label{p_metric} 
\end{equation}
where $H_p(r) = 1 + R^{7-p}/r^{7-p}$, 
$f(r) =  1 - r_0^{7-p}/r^{7-p}$.
Let $\rho = \omega r$. Then the radial part of the 
Laplace equation for a minimally coupled massless scalar
can be written as
\begin{equation}
\frac{d^2 f}{d\rho^2} + 
\left[ a+1 + 
\frac{a \rho^a_0}{(\rho^a - \rho_0^a)} + \frac{a (\omega R)^a (4-a)}{2(\rho^a +(\omega R)^a)}\right]\frac{1}{\rho}
\frac{d f}{d\rho} 
- \frac{\rho^{a-2}l(l+a)}{\rho^a - \rho_0^a} f +
\frac{\rho^a (\rho^a +(\omega R)^a)}{(\rho^a - \rho_0^a)^2} f =0\,,
\label{rho_equation_dp}
\end{equation}
where $a=7-p$, $\rho_0 = \omega r_0$.
Solutions in the ``outer'' region (characterized by
$\rho \ll \rho_0$, $\rho \gg (\omega R)^{a/(a-2)}$) are given by
\begin{equation}
f(r) = \alpha \rho^{-a/2} J_{l+a/2}(\rho ) +\beta 
 \rho^{-a/2} J_{-l-a/2}(\rho ) \,
\end{equation}
for odd $a$ and by
\begin{equation}
f(r) = \alpha \rho^{-a} J_{l+a/2}(\rho ) +\beta 
 \rho^{-a} Y_{l+a/2}(\rho ) \,
\end{equation}
for even $a$.
To study solutions in the ``inner'' region, we
introduce\footnote{The difference in treatment of even and odd $p$ 
reflects our desire to avoid fractional powers of $x$ in the potential.}
 $x= r_0^2/r^2$  for $p$-branes with odd $p$ and  $x=r_0/r$ for
 $p$-branes with even $p$.
Then for the odd $p$-branes the equation (\ref{rho_equation_dp}) becomes
\begin{eqnarray}
&\,& \frac{d^2 f}{d x^2}  + 
\left[ \frac{2-a}{2x} -
 \frac{a x^{a/2}}{2 x (1-x^{a/2})} -\frac{a(4-a)
 x^{a/2}(\omega R)^a/\rho_0^a}{4x(1+x^{a/2}(\omega R)^a/\rho_0^a)}\right]
 \frac{d f}{d x}\nonumber \\
 &+&
\frac{\lambda^2 }{x^{3-a/2}(1-x^{a/2})^2}
\left( 1 + \frac{\rho_0^a}{(\omega R)^a x^{a/2}}\right)f 
- \frac{l(l+a)}{4 x^2 (1-x^{a/2})} f =0\,,
\label{xx_eq_dp_odd}
\end{eqnarray}
where $a = 7-p$ and $\lambda^2 = (\omega R)^a/4\rho_0^{a-2}$.
For  even $p$-branes we get
\begin{eqnarray}
&\,& \frac{d^2 f}{d x^2}  +
\left[ \frac{1-a}{x} +
 \frac{a x^{a-1}}{ (1-x^{a})} - \frac{a(4-a)(\omega R)^a x^a/\rho_0^a}
{2 x (1+(\omega R)^a x^a/\rho_0^a)
}\right]
 \frac{d f}{d x}\nonumber \\
 &+&
\frac{\lambda^2 }{x^{4-a}(1-x^{a})^2}
\left( 1 + \rho_0^a/(\omega R)^a x^{a}\right)f 
- \frac{l(l+a)}{x^2 (1-x^{a})} f =0\,,
\label{xx_eq_dp_even}
\end{eqnarray}
where $a = 7-p$ and $\lambda^2 = (\omega R)^a/\rho_0^{a-2}$.
In the ``inner'' region, $\rho \ll 1$, equations 
(\ref{xx_eq_dp_odd}) and (\ref{xx_eq_dp_even}), correspondingly, become
\begin{equation}
\frac{d^2 f}{d x^2}  + 
\frac{(a^2-6a+4 - (a-2)^2 x^{a/2}}{4x (1-x^{a/2})}
 \frac{d f}{d x} +
\frac{\lambda^2 }{x^{3-a/2}(1-x^{a/2})^2}f
- \frac{l(l+a)}{4 x^2 (1-x^{a/2})} f =0\,.
\label{xx_eq_dp_odd_main}
\end{equation}
\begin{equation}
\frac{d^2 f}{d x^2}  + 
\frac{a^2-6a+2 - (a^2 - 8a +2) x^{a}}{2x (1-x^{a})}
 \frac{d f}{d x}+
\frac{\lambda^2 }{x^{4-a}(1-x^{a})^2}f
- \frac{l(l+a)}{x^2 (1-x^{a})} f =0\,.
\label{xx_eq_dp_even_main}
\end{equation}
Equations (\ref{xx_eq_dp_odd_main}) and (\ref{xx_eq_dp_even_main})
have regular singular points at $x=0,1,\infty$ and a number of others 
on the unit circle. These equations are in this sense similar to the 
black three-brane equation and therefore can be treated by 
the approximation methods used in Section \ref{D3}.

\section{Conclusions}

In this paper we develop a reliable approximation scheme to compute
high- and low-temperature corrections to the absorption cross section 
of a minimally coupled massless scalar in the background of 
near-extremal black branes.  The absorption cross section 
for the non-extremal three-branes is explicitly computed using this scheme,
and compared with the field-theoretic calculation.
Our main results are summarized in Section \ref{summary}.

A number of conceptual and technical points remain to be clarified.
To construct a well-defined scheme based on a hypergeometric
approximation (Section \ref{D3_hyper}) or any modification thereof
would be an important development by itself.  It would allow one to
control exponentially small corrections and therefore quantitatively
compare field theory and gravity results (\ref{f_grav}) - (\ref{f_fth}) 
for the $s$-wave at low
temperature.  
For $l=1$, it would be
interesting to determine the origin of the disagreement between
(\ref{f_l_1_grav}) and (\ref{f_l_1_field}) for $T\ll \omega$.  
In field theory, the spectral densities of correlators with $l>0$ can in
principle be computed exactly. In gravity, corrections to the leading
high-temperature results for higher partial waves (\ref{high_T_odd}), 
(\ref{high_T_even_0}), (\ref{high_T_even_2}) can also be determined.

 It may also be interesting to repeat our
calculations for non-minimally coupled fields.

The approach outlined in Section \ref{M_branes} could be useful
 for studying the world-volume theories on M-branes, as well as  
lower-dimensional QCD (see Section 6.2 of \cite{Aharony:2000ti}
and references therein).

Finally, the results obtained in this paper may prove useful in 
determining transport coefficients of the strongly coupled finite-temperature gauge theory \cite{Son}.

\appendix

\section{Coefficients $A_n$, $B_n$} \label{appendix_a}    
\noindent
Here we provide explicit expressions for the boundary values of
 the first few coefficients  $A_n$, $B_n$ defined by 
(\ref{bar_a}), (\ref{bar_b}).

\noindent
{\bf \em D1-D5 brane system:}
\begin{eqnarray}
A_0 & = & 1\,, \nonumber \\
A_1(0) & = & (32 \nu^3 -12 \nu^2 - 32 \nu + 3)/192\,, \nonumber \\
A_1(1) & = & - (16 \nu^4-8\nu^2 +2)/128\,,  \nonumber \\
A_2(0) & = & (10240 \nu^6 - 44544\nu^5 - 6800 \nu^4 +132480 \nu^3
+17800 \nu^2 - 87936 \nu - 4185)/737280 \,, \nonumber \\
A_2(1) & = & (128 \nu^8 + 2176\nu^6 - 4496\nu^4 +3704\nu^2 -807 )/49152\,.
\end{eqnarray}
Note that ${\cal B}_n$ below is defined as $\lim_{x\rightarrow 1} p(x) B_n(x)$.
\begin{eqnarray}
B_0(0) & = & (\nu^2-1)/6\,,  \nonumber \\
{\cal B}_0(1) & = & (2\nu^2 - 1)/4\,, \nonumber \\
B_1(0) & = & - (116\nu^4 - 345 \nu^2 + 229 )/1920 \,, \nonumber \\
{\cal B}_1(1) & = & - (16\nu^6 +64\nu^4 - 134 \nu^2 +33 )/768\,.
\end{eqnarray}

\noindent
{\bf \em Black three-brane system:}
\begin{eqnarray}
A_0 & = & 1\,, \nonumber \\
A_1(0) & = & 0\,, \nonumber \\
A_1(1) & = & -(\nu^2 -1)^2/8\,, \nonumber \\
A_2(0) & = & (795 + 32768\nu - 420\nu^2 - 40960\nu^3 - 11760\nu^4 + 
      8192\nu^5 + 2880 \nu^6)/40960\,, \nonumber \\
A_2(1) & = &(6801 - 39404\nu^2 + 9648\nu^4 - 
2880\nu^6 + 320\nu^8)/12280  \,, \nonumber \\
A_3(0) & = & 0 \,,\nonumber \\
A_3(1) & = &
    \Biggl(146474829 - 1016514774\nu^2 + 1282758581\nu^4 \nonumber \\
& - &
 553717340\nu^6 + 
        95818800\nu^8 - 5888960\nu^{10} - 24640\nu^{12}\Biggr)/1135411200\,.
\end{eqnarray}
\begin{eqnarray}
B_0(0) & = & 0\,, \nonumber \\
{\cal B}_0(1) & = & (\nu^2 - 1)/2\,, \nonumber \\
B_1(0) & = &  (\nu^4 - 5 \nu^2 + 4)/5\,, \nonumber \\
{\cal B}_1(1) & = & (1344 - 9152\nu^2 + 3520\nu^4 - 320\nu^6)/15360\,, \nonumber \\
B_2(0) & = & 0\,, \nonumber \\
{\cal B}_2(1) & = &
    \Biggl(76248624 - 464811632\nu^2 + 280281408\nu^4 \nonumber \\
 & - &
 60890368\nu^6 + 
        3942400\nu^8 + 78848\nu^{10}\Biggr)/302776320\,.
\end{eqnarray}
Higher coefficients can be rather easily obtained using $\mbox{Mathematica}
^{\mbox{\tiny{TM}}}$ 
but the expression is complicated and will not be presented here.

\section{Asymptotic expansion of
 $|\Gamma (x+ iy)|$ for $|y|\rightarrow\infty$ }\label{appendix_b}    
\noindent
Using the Stirling expansion one can obtain the following formula:
\begin{eqnarray}
|\Gamma (x+ iy)| &=& \sqrt{2\pi} e^{-\pi |y|/2} |y|^{x-1/2} \Bigl\{ 1 +
x(x-1)(2x-1)/12 |y|^2 \nonumber \\
&+& x \Bigl( 20 x^5 - 132 x^4 +245 x^3 -150 x^2 + 5 x +12\Bigr) /1440|y|^4 \nonumber \\
&+& x \Bigl( 280 x^8 - 4284 x^7 + 23046 x^6 - 50337 x^5 + 42735 x^4 \nonumber \\
&-& 2331 x^3- 10801 x^2 +252 x +1440 \Bigr)/362880 |y|^6 + O ( 1/|y|^8 ) \Bigr\}
\end{eqnarray}
Notice that for $x=0, 1/2, 1$ the expression in the curly brackets
 is identically equal to $1$ (no power-like corrections). This is not
 a surprise since we have $|\Gamma (ix)|^2 = \pi/x \sinh{\pi x}$,
 $|\Gamma (1/2 + ix)|^2 = \pi/ \cosh{\pi x}$ and
$|\Gamma (1+ix)|^2 = \pi x/ \sinh{\pi x}$.
\noindent
Another useful fact is the product expansion of
 $\Gamma (x+ iy)$:
\begin{equation}
\Gamma (x+iy) = \xi e^{i\eta}\,,
\end{equation}
where
\begin{equation}
\xi = \Gamma (x)
 \prod_{n=0}^{\infty} \left\{ 1+\frac{y^2}{(x+n)^2}\right\}^{-1/2}\,,
\end{equation}
\begin{equation}
\eta = y \left\{ - \gamma_E + \sum\limits_{n=1}^{\infty}
\left( \frac{1}{n} - \frac{1}{y} \tan^{-1}\frac{y}{x+n-1}\right)\right\}\,.
\end{equation}

\section{Integrals required for a  
thermal field theory calculation of the absorption cross section }\label{appendix_d}    

\noindent
Our approach to evaluating relevant thermal field theory integrals and their
low- and high-temperature asymptotics can be illustrated by the 
following example. We are interested in computing the integral
\begin{equation}
I(b,c) = \int\limits_{0}^{\infty} \frac{\cos{bx}}{x\sinh{c x}}dx = 
-\log{\left( 2 \cosh{\frac{\pi b}{2c}}\right)} = 
-\frac{\pi b}{2c} - \log{ \left( 1+ e^{-\pi b/c}\right)}\,,
\label{d_int}
\end{equation}
where the right-hand side is defined in the sense of analytic continuation.
To calculate (\ref{d_int}) exactly one can either 
take the limit $\alpha \rightarrow 0$ of the expression
(2.5.47.1 (p. 469) of \cite{prudnikov})
\begin{equation}
\int\limits_{0}^{\infty} \frac{x^{\alpha - 1} \cos{bx}}{x\sinh{c x}}dx 
= \frac{\Gamma (\alpha )}{(2c)^{\alpha}}\left\{ \zeta \left[\alpha ,
\frac{1}{2}\left( 1+\frac{ib}{c}\right)  \right] + \zeta \left[\alpha ,\frac{1}{2}\left( 1-\frac{ib}{c}\right)  \right] \right\}\,,
\end{equation}
or evaluate the sum of residues with an appropriate
 prescription for the pole at $x=0$.

\noindent
To obtain a ``low-temperature'' asymptotic,
$c\rightarrow 0$, we can expand $\sinh{cx}$ in series, and then integrate
using  residues. Taking 
 $c\rightarrow 0$ pushes all poles of $\sinh{cx}$ except $x=0$ to
infinity, and thus our integration picks only the ``polynomial'' part
 of the exact result, $-\pi b/2c$. 
Exponentially suppressed terms are not captured. This, however, 
is exactly what we need to compare the field-theoretic absorption
 cross-section with the one obtained from gravity in the form of
 a low-temperature expansion.

\noindent
The other limit, $c\rightarrow \infty$, is more subtle.
In this limit, the poles of  $\sinh{cx}$ merge on the imaginary axis and
one cannot approximate $\sinh{cx}$ by the exponent. One way to proceed 
is to rescale $z = c x$, expand $\cos{bz/c}$ for large $c$, and then integrate.
This produces (in the zeroth-order approximation)
 the right answer, $-\log{ 2 }$, assuming the regularization provided by
the integral
$$
\int\limits_{0}^{\infty} \frac{x^{\alpha -1}}{\sinh{x}}dx =
\frac{2^{\alpha}-1}{2^{\alpha -1}}\Gamma (\alpha )\zeta (\alpha )\,.
$$

\noindent
The spectral density of the two-point function (\ref{the_bos}) for
$l=0$  is given by
\begin{equation}
\rho_0^T (\omega , k) = \frac{\kappa^2}{\pi^2 \omega k}
\left( P(2\pi T, \omega - k) - P(2\pi T, \omega + k)  \right)\,,
\end{equation}
where
$$
P(a,b) = A(a,b) + B(a,b) + I(a,b) + J(a,b) + S(a,b) + T(a,b)\,,
$$
\begin{eqnarray}
A(a,b) &=& \frac{3 a^2}{8} A_{4,2} (a,b) + \frac{a^4}{4}
\left( 1+ \frac{\omega^2}{2 a^2}\right) A_{2,2} (a,b)+
 \frac{3 a^4}{8} A_{2,4} (a,b)\,, \nonumber \\
B(a,b) &=& \frac{3 a^2 \omega }{8} B_{3,2} (a,b) 
+ \frac{a^4\omega }{4}B_{1,2}(a,b)
+ \frac{3 a^4 \omega}{8} B_{1,4} (a,b)\,, \nonumber \\
I(a,b) &=&  \frac{3 a}{8} I_{5,1} (a,b)
 - \frac{a\omega^2}{8}I_{3,1}(a,b) + \frac{3 a^3}{8} I_{3,3} (a,b)
\nonumber \\ &+& \frac{a^5}{8}
\left( 1- \frac{\omega^2}{ a^2}\right) I_{1,3} (a,b)
 + \frac{3 a^5}{8} I_{1,5} (a,b)\,, \nonumber \\
J(a,b) &=& \frac{3 a \omega }{8} J_{4,1} (a,b) 
+ \frac{3 a^3\omega }{8} J_{2,3}(a,b)\,, \nonumber \\
S(a,b) & = & \frac{3 \omega }{8} S_5 - \frac{\omega^3 }{8} S_3\,, \nonumber \\
T(a,b)& = & -\frac{3 \omega^2 }{8} T_4 \,,
\end{eqnarray}
and the integrals involved are defined by
\begin{eqnarray}
A_{m,n}(a,b) &=& \int \frac{\cos{br}\, dr }{r^m\mbox{sinh}^n ar}\,,
 \;\;\;\;\; \; \;\;\;
B_{m,n}(a,b)  =  \int \frac{\sin{br}\, dr }{r^m\sinh^n{ar}}
\\
\nonumber \\
I_{m,n}(a,b) & = & \int \frac{\cosh{ar}\cos{br}\, dr }{r^m\sinh^n ar}
\,, \;\; \; \;\;\; 
J_{m,n}(a,b)  =  \int \frac{\cosh{ar}\sin{br}\, dr }{r^m\sinh^n ar}\,,
\\
\nonumber \\
S_m & = & \int \frac{\sin{br}\, dr}{r^m}\,, \;\;\;\hspace{15mm}
T_m  = \int \frac{\cos{br}\, dr}{r^m} \,.
\end{eqnarray}
Calculation gives
\begin{eqnarray}
I_{5,1}(a,b) &=& 
\frac{\pi b\,\left( 8\,a^4 + 20\,a^2\,b^2 - 3\,b^4 \right) \,
      }{360\,a} + \frac{2\,a^4\,p_5}{\pi^4}\,.\\
\nonumber \\
I_{3,1}(a,b) &=& \frac{b\,\left( -2\,a^2 + b^2 \right) \,\pi }{6\,a} - 
  \frac{2\,a^2\, p_3}{\pi^2}\,, \\
\nonumber \\
I_{3,3}(a,b) &=&
-\frac{\pi b (b^4-8a^4)}{120 a^3} +
\frac{12 a^2 p_5}{\pi^4}
+\frac{6 a b p_4}{\pi^3}
+\frac{ b^2 p_3}{\pi^2}\,,\\
\nonumber
\\
I_{1,3}(a,b) &=&
\frac{\pi b^3}{6 a^3} -
\frac{2 p_3}{\pi^2}
-\frac{2  b p_2}{\pi a} 
-\frac{ b^2 p_1}{a^2}\,,\\
\nonumber
\\
I_{1,5}(a,b) &=&
- \frac{\pi b^3 (20 a^2 + 3 b^2)}{360 a^5} 
+\frac{2 p_5}{\pi^4}
+\frac{2 b  p_4}{\pi^3 a}
+\left( \frac{2}{3\pi^2}+\frac{b^2}{\pi^2 a^2}  \right) p_3 \nonumber \\
&+&\left( \frac{b^3}{3\pi a^3}+\frac{2 b}{3\pi a}\right) p_2
+\left( \frac{b^4}{12 a^4}+\frac{b^2}{3 a^2}\right) p_1\,,\\
\nonumber
\\
J_{2,3}(a,b) &=&
- \frac{\pi a}{15}+\frac{\pi b^4}{24 a^3}  
+\frac{6 a p_4}{\pi^3}
+\frac{b  p_3}{\pi^2}+\frac{b^2  p_2}{\pi a}\,,\\
\nonumber
\\
J_{4,1}(a,b) &=&
- \frac{\pi a^3 }{45}- \frac{\pi a b^2}{6}+ \frac{\pi b^4}{24 a} 
+\frac{2 a^3 p_4}{\pi^3}\,,\\
\nonumber
\\
A_{4,2}(a,b) &=&
- \frac{\pi b(24 a^4 + 20 a^2b^2 +3 b^4) }{360 a^2}  
-\frac{8 a^3 p_5}{\pi^4}-\frac{2 a^2 b p_4}{\pi^3}\,,\\
\nonumber
\\
A_{2,2}(a,b) &=&
 \frac{\pi b(2 a^2 +b^2)}{6 a^2}  + \frac{4 a p_3}{\pi^2}+
\frac{2 b p_2}{\pi}\,,\\
\nonumber
\\
A_{2,4}(a,b) &=&
- \frac{\pi b(88 a^4 +40 a^2b^2+3 b^4)}{360 a^4}-\frac{8 a p_5}{\pi^4}
-\frac{6 b  p_4}{\pi^3}-\left( \frac{8 a}{3\pi^2}
 +  \frac{3 b^2}{\pi^2 a} \right) p_3 \nonumber \\
&-&
\left( \frac{b^3 }{3 a^2 \pi}+\frac{4 b}{3\pi}\right) p_2
\,,\\
\nonumber
\\
B_{1,2}(a,b) &=&
- \frac{\pi ( 2 a^2 + 3 b^2)}{6 a^2}
+\frac{2 p_2}{\pi}
+\frac{2 b p_1}{a}
\,,\\
\nonumber
\\
B_{1,4}(a,b) &=&
 \frac{\pi (88 a^4 +120 a^2 b^2+15 b^4)}{360 a^4}
- \frac{2  p_4}{\pi^3}- \frac{2 b  p_3}{\pi^2 a}
-2 \left( \frac{2 }{3 \pi } +  \frac{b^2}{2\pi a^2}\right) p_2 \nonumber \\
&-& 2 \left( \frac{2 b}{3 a}
+  \frac{b^3}{6 a^3}\right) p_1
\,,\\
\nonumber
\\
B_{3,2}(a,b) &=&
 \frac{\pi ( 8 a^4 + 20 a^2 b^2 + 5 b^4)}{120 a^2}
-\frac{6 a^2 p_4}{\pi^3}
-\frac{2 a b p_3}{\pi^2}\,,
\end{eqnarray}
where $p_n(a,b)= \mbox{Li}_n [\exp{(-\pi b/a)}]$ is a polylogarithm
 \cite{prudnikov}. We also have $S_3 = -\pi b^2/2$,  $S_5 = \pi b^4/24$,
 $T_5 = \pi b^3/6$.

\vskip .2in \noindent \textbf{Acknowledgments}\vskip .1in \noindent
We would like to thank P.~A.~Grassi, M.~Porrati and D.T.~Son for useful 
discussions and comments, and B.~Bajc and F.~W.~J.~Olver
for correspondence.
G.P. is supported in part by the Fondazione A. Della Riccia.

\end{document}